\documentstyle[aaspp4,12pt]{article}
\newcommand{\etal}{{\em et~al.}}

\begin{document}
 
\title{The Distribution of Collisionally Induced Blue Stragglers in
the Colour-Magnitude Diagram}
\author{Alison Sills} 
\author{Charles D. Bailyn \altaffilmark{1}}
\altaffiltext{1}{National Young Investigator}

\affil{Department of Astronomy,
Yale University, P.O. Box 208101, New Haven, CT, 06520-8101}

\begin{abstract}

A primary production mechanism for blue stragglers in globular
clusters is thought to be collisionally-induced mergers, perhaps
mediated by dynamical encounters involving binary stars.  We model the
formation and evolution of such blue stragglers, and produce
theoretical distributions of them in the colour-magnitude diagram.  We
use a crude representation of cluster dynamics and detailed
binary-single star encounter simulations to produce cross sections and
rates for a variety of collisions. The results of the collisions are
determined based on SPH simulations of realistic star models.  The
evolution of the collision products are then followed in detail using
the Yale stellar evolution code.  We present our results in the form
of distributions in the observed colour-magnitude diagram.

We use our results to explore the effects of a variety of input
assumptions on the number and kind of blue stragglers created by
collisions.  In particular, we describe the changes in the blue
straggler distribution that result from using realistic collision
products rather than the ``fully-mixed'' assumption, and from changes
in assumptions about the number and orbital period distribution of the
primordial binary population.  We then apply our models to existing
data from the core of M3, where the large blue straggler population is
thought to be dominated by collision products.  We find that we have
difficulty successfully modeling the observed blue stragglers under a
single consistent set of assumptions.  However, if 3 particularly
bright blue stragglers are considered to be part of a different
observed population (as has been previously suggested for other
clusters) and left out of the population, the remainder can be
successfully modeled using realistic encounter products and assuming a
20\% binary fraction with plausible period distribution.  Finally, we
suggest a variety of routes towards a more comprehensive understanding
of the blue straggler phenomenon.
\end{abstract}


\section{Introduction}

Globular clusters are important astrophysical laboratories for two
different processes: stellar evolution and stellar dynamics. A
population of stars with the same age, composition and distance has
been an ideal place in which to test theories of stellar evolution,
beginning with the basic evolutionary tracks and now including such
details of stellar structure and evolution as diffusion of heavy
elements or convective overshoot (e.g. Chaboyer \etal\ 1992, Brocato
\etal\ 1997). A globular cluster can also be considered to be a dense
system of $\sim 10^5$ point masses which interact only under the
influence of their mutual gravity. These simple systems are obvious
choices in which to test N-body calculations and other dynamical
simulations (Dull \etal\ 1997, Elson \etal\ 1987, Meylan \& Heggie
1997 and references therein).

It is becoming apparent, however, that we can no longer study these
two astrophysical processes independently if we are to completely
understand globular clusters.  Stellar evolution tells us that
contrary to the simple dynamic assumptions, stars are not point masses
with infinite lifetimes, but instead have finite radii and have
lifetimes which are shorter for more massive stars.  These properties
can have a significant impact on the dynamical processes. For example,
a result of equipartition of energy in the cluster is mass
segregation, which causes the more massive stars to sink toward the
centre of the cluster (Lightman \& Shapiro 1978). These massive stars
have shorter lifetimes than their less massive counterparts, and will
evolve into stellar remnants more quickly. These remnants tend to have
a smaller mass than their parent star, and so the distribution of
masses in the core of the cluster will change as the stars
evolve. This change in the mass function will affect the rate of core
collapse and other dynamical processes (Chernoff \& Weinberg 1990).

The presence of stars with finite radii allows for the possibility of
physical stellar collisions or tidal interactions. These collisions
will change the energy budget in the cluster, which can influence the
timescales on which core collapse, mass segregation and other
dynamical processes occur (Spitzer 1987). The collision products may
be objects which cannot be explained by standard stellar evolution
theory, such as blue stragglers, low mass X-ray binaries and
millisecond pulsars (see Bailyn 1995 for a review). Since these stars
have very different masses and lifetimes than normal globular cluster
members, their presence can affect the mass function of a region of
the cluster, which can then modify the timescales for dynamical
processes.  Therefore, the study of stellar collision products can
lead to an understanding of the interactions between stellar evolution
and stellar dynamics in globular clusters.

One possible class of collision product to study is blue straggler
stars.  Blue stragglers are stars which are bluer and brighter than
the main sequence turnoff. They have the same surface gravity values
as main sequence stars (Rodgers \& Roberts 1995, Shara \etal\ 1997),
and therefore must have been born more recently that the majority of
stars in the cluster.  Blue stragglers are thought to be created by
the merger of two normal globular cluster stars, either through the
merger of the two components of a binary star system, or through
direct stellar collisions (Stryker 1993).  Both creation mechanisms
are expected to occur, and which mechanism is dominant depends on the
binary population and the density of the environment.

Recent studies of the creation and subsequent evolution of collisional
blue stragglers (Bailyn \& Pinsonneault 1995, Sills \etal\ 1995,
Ouellette \& Pritchet 1996, 1998, Sandquist \etal\ 1997) used the
results of SPH simulations of stellar collisions to predict the
initial structure and composition of blue stragglers. A number of
different cases were then evolved using a stellar evolution code.  The
results were presented as a series of evolutionary tracks, or simple
assumptions were made about the collision rate and luminosity
functions were calculated (Bailyn \& Pinsonneault 1995). In this
paper, we extend these approaches, and combine the evolutionary tracks
with detailed binary-single star encounter cross sections and crude
cluster dynamics to model the distribution of blue stragglers in the
colour-magnitude diagram of globular clusters.

In section 2, we present a method for calculating the total number and
distribution of collisional blue stragglers in the colour-magnitude
diagram. In section 3, we discuss our input assumptions, and in section
4 we produce theoretical colour-magnitude diagrams of blue stragglers
for a variety of these assumptions. In section 5, we apply this
method to the globular cluster M3 as a specific example.

\section{Method}

We wish to determine the distribution of blue stragglers in a
colour-magnitude diagram. In order to do that, we must calculate the
number of stars in each (colour, absolute magnitude) box as follows:
\begin{equation}
N(C, M) = \sum_{i} \sigma_i f_i t_i(C,M) 
\end{equation} 
summed over all collision products $i$, where $\sigma_i$ is the cross
section for the $i$th collision product, a function of the parents of
the collision and the outcome; $f_i$ is the flux of the parents of the
collision which results in a particular outcome $i$, and $t_i(C,M)$ is
the time spent in a box of $(C + dC, M + dM)$ on the evolutionary track of
the collision product. Each of these three components are discussed in
detail in the following sections.

\subsection{The Cross Sections}

The cross sections for collisions between single stars is
(e.g. Spitzer 1987)
\begin{equation}
\sigma_{ss}=\pi (r_1+r_2)^2 \left(1+\frac{2G(m_1+m_2)}{(r_1+r_2)v_{rel}^2}\right)
\end{equation}
where $m_1$ and $m_2$ are the masses of the two colliding stars, $r_1$
and $r_2$ are their radii, and $v_{rel}$ is the relative velocity of
the two stars at infinity.

Direct stellar collisions can also be mediated by binary
systems. Since binary systems have a larger effective area than single
stars (proportional to the semi-major axis rather than the stellar
radius), interactions between single stars and binaries are more
likely than between two single stars for a given density of the
environment (Hills 1975, Heggie 1975, Leonard 1989). When a binary and
a single star interact, many outcomes are possible. In general, the
third star perturbs the binary system for a time. If that perturbation
is large enough, the three stars undergo a resonance encounter and
form an unstable triple system with a very complicated orbit (Hut \&
Bahcall 1983). During this period, two, or even all three of the
stars, can pass close enough to each other to physically collide.

We used a program based on STARLAB (McMillan 1996) for calculating
cross sections for collisions between binary and single stars.  This
program is an automated Monte Carlo routine which chooses binary
systems and single stars from some specified distribution, and then
fires the single star at the binary system with a range of incoming
angles. The results of each interaction are tallied, and a cross
section is determined as a function of the binary system, the single
star, and the resultant collision product (or lack thereof).  
Collisions between two stars and those between all three stars (triple
collisions) are taken into account.

\subsection{The Fluxes}

The fluxes of the collision parents depend on many parameters, some
relevant to the cluster as a whole, and some dependent on the specific
star or binary system involved.  Fluxes are of the form \(f=\rho v\)
where $\rho$ is the density of the objects involved, and $v$ is their
relative velocity. A full exploration of the effects of the range of
relative velocities and the variation of density with cluster radius
expected in real physical systems requires a more complete dynamical
model than we can explore in this paper.  For simplicity, we have
assumed that all objects in the cluster have a relative velocity equal
to the velocity dispersion of the cluster. The overall density of the
region of interest (the core of the cluster) was assumed to be
constant and equal to the central density of the cluster. We need to
then determine what fraction of that density is in the form of the
particular parent stars and binary systems we wish to consider. The
density of collision parents is simply the total density multiplied by
the probability of choosing such systems from the total distribution
of systems. That probability depends on the distribution of stars by
mass, the binary fraction in the region of the cluster, and the period
distribution of the binary systems.

We have assumed a mass function of the form
\begin{equation}
\frac{dN}{dM}=M^{-(1+x)}
\end{equation}
so that $x=1.35$ for a Salpeter mass distribution. Therefore, the probability
of choosing a particular mass $m$ within $dm$ from this distribution is
\begin{equation}
P(m,m+dm)=\frac{m^{-x}-(m+dm)^{-x}}{x N_{tot}}
\end{equation}
where 
\begin{equation}
N_{tot}=\int_{M_{low}}^{M_{high}} M^{-(1+x)} dM
\end{equation}

For collisions between two single stars, the probability of choosing the two
stars in question from the global mass function is simply the product of 
the individual probabilities of choosing those stars, and so the flux
of parents for a single star -- single star collision is 
\begin{equation}
f_{ss}=P(m_1) P(m_2) \rho v
\end{equation}
where $P(m)$ is defined above.  

When a single star collides with a binary system, there are more
parameters to consider.  We must specify the masses of the binary
system, $m_1$ and $m_2$, the period of the binary $P$ and its
eccentricity $e$.  In order to specify how likely it is to have a
particular combination of $m_1$ and $m_2$ in a binary system, we must
make some assumption about how the binary system was formed.  In
general, we assume that the binary components were randomly selected
from a particular mass function which is not necessarily the same as
the mass function of the single stars. Then we must include the
distribution of binary systems by total mass in the cluster. For this,
we again use a mass function of the same form as in equation 3 above.

Thus, there are three mass functions relevant to the problem: one
describing the distribution of single stars in the core of the
cluster, one describing the distribution of binary stars in the
cluster, and one describing the distribution of the components of the
binary system. In principle, these three mass functions can be
different, but in practice we have assumed that the single stars and
the binary systems are currently distributed in the cluster with the
same mass function. This is a reasonable assumption since mass
segregation causes the more massive objects to sink to the centre of
the cluster, resulting in a mass function heavily skewed toward
massive stars and binaries. This should result in similar mass
functions regardless of initial distribution. When the binary systems
formed, however, their components were presumably drawn from the same
initial mass function as the single stars, which will include fewer
massive stars than the current mass function. Note that we are
assuming that the binary systems are primordial. We must also include
the binary fraction, which we have defined to be the fraction of
systems which are binary \footnote{Some authors define the binary
fraction to be the fraction of stars in the cluster which are in a
binary system. There is a simple transformation from one definition to
the other.}.

Therefore, the flux of parents of a collision between a single star
and a binary system is 
\begin{equation}
f_{bs}=P(m_1) P(m_2) P(P) P(m_1+m_2) f_{bin} P(m_3) \rho v
\end{equation}
where $f_{bin}$ is the binary fraction.

\subsection{The Evolutionary Tracks}

The evolutionary tracks were calculated using the Yale Rotating
Evolution Code (YREC) in its non-rotating configuration. We used the
latest OPAL opacities (Iglesias \& Rogers 1996) for the interior of
the star down to temperatures of $\log T=4$. For lower temperatures,
we used the low-temperature opacities of Alexander \& Ferguson
(1994). For a detailed description of the other physics included in
this stellar evolution code, see Guenther \etal\  (1992).

We have calculated two sets of evolutionary tracks for blue
stragglers. For the first set, we assumed that stellar collisions
between globular cluster stars are violent events which result in
fully mixed collision products (Benz \& Hills 1987). These models are
calculated in the following way. Each parent star in the collision is
evolved to 15 Gyr and the total mass of helium in each star is
calculated. The two (or three, for a triple collision) helium masses
are summed, and the collision product is assumed to have lost 5\% of
its total mass during the collision (10\% for a triple collision). The
new helium content, Y, of the collision product is the total helium
mass of the parent stars divided by the total mass of the collision
product. The stars were then evolved from the zero age main sequence
up the giant branch.

A more realistic set of evolutionary tracks can be created by using
the results of smoothed particle hydrodynamics (SPH) simulations of
stellar collisions as starting models for stellar evolution
calculations. This process is described in detail in Sills {\it et
al.} (1997). The tracks presented in that work (and in Sandquist
\etal\ 1997) resulted from SPH simulations of head-on collisions
between parent stars which were assumed to be polytropes. However,
polytropes are not reasonable approximations to evolved main sequence
stars. Instead, stellar models taken from evolutionary codes should be
used as the initial conditions for SPH simulations (Sills \& Lombardi
1997). Therefore, rather than redoing the SPH simulations, we used an
interesting feature of the SPH algorithm to determine the structure of
the collision products without actually performing the SPH
simulations. The results of the SPH simulations can be predicted by
sorting the fluid from the parent stars in order of increasing
entropy.

We wish to reproduce as exactly as possible the stellar collision
products as calculated by the smoothed particle hydrodynamics code. In
order to accomplish this, we need to understand what the SPH code
does, so that we can mimic it.  A stellar collision calculation has
reached hydrodynamic stability and is terminated when the Ledoux
criterion is satisfied. This criterion can be written \( \frac{dA}{dr}
\geq 0 \), where $A=P/\rho^{\Gamma}$ is directly related to the
entropy of the gas. The quantity $A$ can only be changed by shock
heating. Therefore, in the absence of shocks, the collision product is
determined by ordering the particles of the two parent stars in order
of increasing $A$. In globular clusters, the collisions between two
stars are quite slow, and most of the shock heating is limited to the
outermost few percent of the two stars. The interior structure of the
collision product is not affected.  The stability criterion is only
valid for non-rotating collision products, so we are limited to the
head-on collisions if we use this method.
 
Figure 1 shows the profiles of $A$ and $Y$ vs. mass fraction for a
collision product calculated using the SPH code of Lombardi {\it et
al.} (1996), and the profiles determined by the algorithm given above
for a head-on collision between a $0.8 M_{\odot}$ star and a $0.4
M_{\odot}$ star. The stars are assumed to be ideal gases, with
$\Gamma=5/3$.  After determining the $A$ vs. mass and $Y$ vs. mass
profiles, we then create a stellar model from those profiles as
described in Sills \etal\ (1997).  The final approximation
necessary to make this product more closely resemble the results of
the SPH calculations is to reduce the star's total mass by 5\%, since
the collision does not result in all the mass being bound to the
merger remnant. This mass is removed by reducing the mass in each
shell.

As can been seen in figure 2, this procedure works well for a collision
between a $0.8 M_{\odot}$ star and a $0.4 M_{\odot}$ star. The solid
line is the track which results if we use our algorithm for mimicking
the SPH calculation to create a starting model for YREC. Since we
neglected shock heating, this star has not increased its energy
content as much and begins its life much closer to the main
sequence. However, the chemical profile is quite similar to that of
the SPH collision product, and so the evolution of the star follows
the same path through the HR diagram after thermal equilibrium is
reached. The thermal relaxation phase of evolution is quite short
(about $10^6$ years in this case) and the timescales of the two tracks
are essentially the same from the `zero-age' main sequence as well,
indicating that we can effectively determine the future evolution of
collision products simply by knowing the pressure, density and
chemical profiles of the parent stars.

From the evolutionary tracks, we can determine $t_i(T_{eff},L)$, the
time spent in each box in the HR diagram for a particular collision
product. This time is combined with the flux and cross section to
determine that collision's contribution to the number of blue
stragglers at each position in the HR diagram.

\subsection{Conversion to Colour-Magnitude System}

It has been standard practice to convert theoretical evolutionary
tracks to ground-based UBVRI magnitudes in order to more easily
compare theory and observation. Since globular clusters are so dense
and therefore so crowded, it is often necessary to use the Hubble Space
Telescope to observe the cores of clusters. If we wish to compare our
theoretical predictions with HST observations, we must convert our
evolutionary tracks into HST system magnitudes. We calculated a
transformation table with colours as a function of temperature,
gravity and metallicity for the HST filters most like the Johnson
ground-based UBVRI system, with one addition (F255W, F336W, F439W,
F555W, F675W, F814W). This table was calculated from Kurucz
atmospheres, using a similar technique to the one Yi \etal\ (1995) used
for the FOC ultraviolet filters, and is available by request from the
authors.

\section{Choice of Parameters}

To explore the effects of different input assumptions regarding the
binary star population, we assumed a set of values for the velocity
dispersion and central density of a `standard' globular cluster.  We
assumed a velocity dispersion of 10 km/s, which is a typical value for
globular clusters. We assumed the density in the core to be $8.75
\times 10^3$ stars/pc$^3$, or or about $7 \times 10^3
M_{\odot}$/pc$^3$, an appropriate value for a reasonably dense cluster
core prior to core collapse.

We calculated cross sections for collisions between single stars using
equation 2. We used the program based on STARLAB to calculate cross
sections for collisions between binary and single stars. For this
program, we assumed a relative velocity of 10 km/s, and a flat mass
function ($x = -1$) in order to evenly sample the mass ranges for both
the binary and single stars. We then weighted the results
appropriately to determine the results for other mass functions. Our
upper and lower mass limits were $0.4 M_{\odot}$ and $0.8 M_{\odot}$
respectively.  We calculated cross sections for semi-major axes from
0.01 AU to 30 AU. The lower limit corresponds to stars which are not
quite in contact, and the upper limit is approximately the maximum
separation of a hard binary system (one with a binding energy greater
than the local stellar thermal energy). Binaries with larger
semi-major axes are soft binary systems, and are destroyed in globular
clusters within a short time. We assumed a thermal eccentricity
distribution ($f(e)=2e$) for the binary systems, except for the close
systems. Binary systems with semi-major axes less than 0.1 A.U. were
assumed to be circular, due to tidal circularization (Mathieu \& Mazeh
1988).

As described in section 2.2, we use a number of different mass
functions in this work. We therefore adopt a variety of exponents for
the current distribution of stars and binary systems, ranging from
Salpeter ($x=1.35$) to $x=-2$, which is strongly weighted towards
massive stars. 

The binary fraction in globular clusters is not well known. Therefore,
we have considered the entire range of binary fractions, from 0 to
100\%.  If we assume a flat period distribution in $\log P$ for the
binary stars, as suggested by Yan \& Mateo (1994) for M71, the
probability of choosing a particular binary period $\log p$ within $d
\log p$ is \(P(\log p)=d \log p/(\log p_{\rm max}-\log p_{\rm min}
)\). We have chosen our maximum and minimum periods to be $\sim 3
\times 10^{4}$ and $\sim 2 \times 10^{8}$ seconds respectively, which
roughly correspond to the maximum and minimum semi-major axes used to
determine the cross sections.  The other option we have explored is
the Duquennoy \& Mayor (1991) period distribution for G dwarfs in the
solar neighbourhood, which is weighted towards longer period
distributions. To explore the efferent effects of long and short
period binaries, we have also considered two artificial period
distributions, which are flat in $\log P$ but have either a high or
low period cutoff at $ P = 10^6$ seconds.

After the collisions have occurred, the stars evolve. As described
above, we have two sets of evolutionary tracks to choose from: fully
mixed, and `SPH' tracks. The fully mixed tracks provide an upper limit
on the amount of mixing which can occur during the formation of the
blue stragglers. Table 1 gives a list of parameter combinations for
which we have calculated blue straggler distributions, as well as the
total number of blue stragglers predicted to exist in the region of
interest.

\section{Results}

The resulting distributions of blue stragglers in the colour-magnitude
diagram are shown in figures 3 through 7.  Since the data are
presented in the somewhat unfamiliar HST ultraviolet filters, figure 3
presents the landmarks necessary to interpret the following
figures. We have plotted the zero-age main sequence (ZAMS) for M3, as
well as an evolutionary track of a turnoff mass star. The fiducial
sequence of the cluster follows this evolutionary track quite well,
and all blue stragglers should lie between the ZAMS and the
evolutionary track, brighter than the turnoff. The crosses are the
data points from Ferraro \etal\ 1997. The large box surrounding those
data points is the region shown in the following plots, and the small
box to the lower right is 0.16 magnitudes on a side, which shows the
resolution of the theoretical contours.

In figures 4 through 7, the grey scale gives the relative density of
blue stragglers at each position in the colour-magnitude diagram, with
5 density intervals at levels of 1, 5, 10, 50, and 100 stars per box
0.16 magnitudes on a side.  These subsequent figures demonstrate the
effects of the different parameters on the shape of the distributions
and on the total number of blue stragglers produced, as discussed in
the following section.

\subsection{Effects of Parameters}

The overall shape of the distribution is most sensitive to the choice
of evolutionary tracks. The blue stragglers which were fully mixed by
the collision produce a distribution which remains quite close to the
ZAMS, since all stars begin their life on the ZAMS. The distribution
also reaches to brighter stars, because these stars have higher helium
content than normal for stars in the cluster. The helium which was
created in the cores of the collision parents has been spread
throughout the star, creating a collision product which is bluer and
brighter than might otherwise be expected. The results of the SPH
simulations produce distributions which are more spread out in
colour than the fully mixed distributions, and does not extend to the
high brightness of the mixed models.  The distributions which result
from these two kinds of evolutionary tracks are displayed in figure 4,
and compared again in figure 5.

The binary fraction determines how many binary star systems are
available for collision. If the binary fraction is higher, more
collisions will occur, increasing the total number of blue
stragglers. Figure 5 compares distributions for binary fractions of
0\%, 20\% and 100\%. For this figure, the distributions have been
normalized to 100 blue stragglers each to emphasize the change in
shape rather than the change in total number of blue stragglers. The
total numbers produced are recorded in table 1. While the total
number of blue stragglers increases dramatically as the binary
fraction increases, the underlying shape of the distribution does not
change noticeably once the binary faction rises above zero. Since the cross
section for collision for binary stars is so much larger than that
between single stars, the binary-single contributions quickly
overpower the contribution of the single-single collisions.

The period distribution of the binary systems affects the kinds of
collisions that occur. If there are more short-period systems, triple
collisions are more likely and we would expect to see more massive
blue stragglers than low mass objects. Figure 6 shows blue straggler
distributions calculated using a period distribution which is flat in
$\log P$, the Duquennoy \& Mayor (DM) period distribution, and two
flat period distributions which have been artificially restricted to
$\log P \leq 6$ and $\log P > 6$ for periods in seconds. These
distributions have not been normalized and represent the actual number
of stars created in each case. The overall differences in shape
between the flat period distribution and the DM period distribution
are quite small. The major difference between the two distributions of
blue stragglers is that the DM period distribution produces more total
blue stragglers than the flat distribution. The longer-period binary
systems are more likely to interact with a passing single star, and so
the cross sections for interaction are larger, increasing the
contribution to the total distribution. The same effect is seen more
clearly by comparing the two distributions with the high and low
period cutoffs. The shape of the distribution, when normalized, is
almost the same, but the long period group produces many more blue
stragglers. The masses of these binary systems are not constrained by
choosing a particular period distribution, and so the shape of the
distributions does not change significantly.

As described in section 2.2, there are two mass functions which
describe the populations involved in the collisions, namely the
current mass function in the core, and the mass function from which
the binary components are drawn. The nine distributions in figure 7
present the results of some possible combinations of these two mass
functions. These distributions have been normalized to 100 total
stars. The exponent of the current mass function decreases from 1.35
to -2 from left to right, and the exponent of the binary mass function
decreases from 1.35 to -2 from top to bottom. Mass functions with
negative exponents are weighted towards massive stars. As expected,
the greater the fraction of massive stars, the brighter the
distribution becomes. This is true as either of the mass functions is
changed, but the mass function of the binary components causes more
change in the shape of the distribution than the current mass
function. This suggests that the distribution of stars in the binary
systems is more important than the population of the single stars in
determining the total number and distribution of blue stragglers in
the colour-magnitude diagram.  

We have used a Kolmogorov-Smirnoff (KS) test in two dimensions using
the algorithm given by Press \etal\ 1992 to quantify the differences
between the theoretical blue straggler distributions.  This test is
similar to the one dimensional KS test in that it gives the
probability that the two distributions are drawn from the same
distribution. While the two-dimensional test is not as robust as the
one-dimensional test, it can indicate if two distributions are
significantly different from each other. We compared a number of
different distributions to our `standard' distribution, case J. This
model uses the SPH tracks and has a binary fraction of 20\%, a flat
binary period distribution, a current mass function with x=-2, and a
binary component mass function of x=1.35. The KS statistic $d$ and the
probability that the two distributions are drawn from the same
underlying distribution are presented in table 2. The total number of
points in each distribution is given by the model predictions, and
therefore the KS statistics are relevant for comparisons between two
theoretical clusters. These statistics support the previous claims
that the evolutionary tracks and binary mass function are more
important to the overall shape of the theoretical distributions than
the period distribution and current mass function, given the predicted
total number of blue stragglers.

\section{Comparison with Observations: M3}

As a specific example, we have modeled the distribution of blue
stragglers in the globular cluster M3. M3 is interesting for a number
of reasons. It has the largest sample of blue stragglers of any
globular cluster, which means that it is possible to discuss
distributions of blue stragglers with some statistical
significance. M3 also has a very odd radial distribution of blue
stragglers which has not yet been seen in any other cluster, possibly
because of small number statistics. Rather than being spread evenly
throughout the cluster or simply centrally concentrated, the blue
stragglers appear to be concentrated in the core of the cluster, and
in its outskirts, with a deficit of blue stragglers between 100 and
200 arc seconds from the cluster centre (Ferraro \etal\ 1993, Bolte
\etal\ 1993). This may suggest two different blue straggler creation
mechanisms at work. Bailyn \& Pinsonneault (1995) suggested blue
stragglers in the core of M3 are created by stellar collisions, and
those in the sparser outer regions of the cluster are created by the
merger of a binary system.  On the other hand, Sigurdsson {\it et al.}
(1994) suggest that all the blue stragglers in M3 are created by
stellar collisions mediated by binary stars, and the outer blue
stragglers were kicked out of the centre of the cluster by the recoil
of the collision. In either case, the central blue stragglers in M3
are expected to have been created by stellar collisions.

In order to compare our theoretical results with observational data,
we require data with a number of qualities. The data must be complete,
both in magnitude and in position in the cluster. If the data are
complete, we can be sure that their distribution is the distribution
we must reproduce, and we can be sure that the total numbers of blue
stragglers is accurate. We also need a large sample of blue stragglers
so that any comparison between data and theory can be studied in a
statistical sense, and we are not limited by small numbers.
Observations in the cluster M3 come closest to satisfying these
criteria of any currently available data set.

Many groups (e.g. Ferraro \etal\ 1993, Bolte \etal\ 1993) have
observed blue stragglers outside the core of M3, and many of those
data sets have good photometry and are complete. We have chosen to
restrict our investigation to the core of the cluster since that is
the region in which we expect stellar collisions to dominate the blue
straggler population. Since we are assuming that all blue stragglers
are created through stellar collisions, we wish to restrict ourselves
to a dense environment where this approximation is more appropriate.

We use the HST photometry of Ferraro \etal\ (1997) of the core of M3
as the data to which we compare our theoretical distributions.  These
observations cover the core of the cluster, and were taken in HST
ultraviolet filters specifically to study the hot stars (including the
blue stragglers) in the cluster. This data set is reported to be
complete down to $m_{255} = 19.0$ for the entire area of the cluster
covered, which includes the core. There are 72 blue stragglers in the
inner 100'' of the cluster with $m_{255}$ brighter than 19 in this
sample. The photometric errors in this data are fairly large, on the
order of 0.15 magnitudes in both colour and magnitude at the main
sequence turnoff.

\subsection{Statistical Comparison of Theoretical Distributions to Data}

The two distributions which used different evolutionary tracks (shown
in figure 4) have widths which are different in a statistically
significant way. To demonstrate this, we took the average distance
from the ZAMS line in magnitude bins of size 0.2 mag from M=3.9 to
M=1.3. We then performed a two-sided Kolmogorov-Smirnoff (KS) test
which determines if the two distributions were drawn from the same
underlying distribution. We determined that the two distributions were
drawn from the same distributions with a probability of 0.01\%. We
then compared each theoretical distribution to the observational data
in the same way. If all data points are included in the KS test, the
SPH tracks have a 3.7\% chance of being drawn from the same
distribution as the observational data, and the fully mixed tracks
have a 26\% chance of being drawn from the same distribution. Neither
of these results are conclusive, but the indication is that the fully
mixed tracks better represent the data. However, if we exclude the
three brightest stars from the data set and study the region from
M=3.9 to M=2.1, we get significantly different results. Now, the SPH
tracks have a 67\% probability of being drawn from the same
distribution, and the fully mixed tracks have a 3\% chance. This
result is only valid at the 1-$\sigma$ level, but it is strongly
indicative that the fainter portion of the data is better modeled with
the SPH tracks, and the brightest three blue stragglers in M3 may be
different from the rest of the population.

We also created luminosity functions of our theoretical distributions
by binning our distributions, and similarly we created temperature
functions. These one-dimensional distributions can be compared to the
data using a KS test. The data was binned in bins of 0.2 mag in
luminosity, and 0.1 mag in temperature and then compared to the
theory. The KS statistic $d$ and the related probability that the data
is well-matched by the theory are given in columns 2 through 5 of
table 2. In general, KS tests only distinguish between distributions
which are not drawn from the same underlying distribution. Therefore,
any distribution which has a probability of being the same as the
observational distribution within about 50\% is considered to be a
good match to the observations. Many of the theoretical distributions
have KS probabilities which are greater than 50\% for the luminosity
function, the temperature function, or both.  However, we can see that
many of the distributions which have decent KS statistics for either
the luminosity function or the temperature function do not fit the
data well. Therefore, we conclude that it is necessary to compare both
axes of the colour-magnitude diagram simultaneously, rather than
simply considering the luminosity function (Bailyn \& Pinsonneault
1995) or the distance from the ZAMS (Ouellette \& Pritchet 1996) alone.

To demonstrate the need for photometry with small observational
errors, we have convolved one distribution (case J) with a Gaussian of
width 0.15 magnitudes in both magnitude and colour. The observational
errors on the Ferraro \etal\ (1997) data set are quoted to be about
0.15 magnitudes at the turnoff. As can be seen in figure 8, errors of
this size mask any structure in the theoretical distributions, and
only general statements about height and width can be made. It is
interesting to note, however, that even when the observational errors
are taken into account, the brightest three blue stragglers in M3 are
not modeled by the distribution. This result is even more striking
when we consider that these bright blue stragglers are at least 3
magnitudes brighter than the cluster turnoff, and therefore we expect
that the photometric errors on these stars are much smaller than 0.15
magnitudes (as is demonstrated by the lack of scatter in the
horizontal branch in the same data set). If such small observational
errors are assumed, the probability that any of our theoretical
distributions can fit the observed stars, including the brightest
three, is greatly reduced.

\section{Discussion}

\subsection{Validity of Assumptions}

In order to study one component of the interaction of stellar dynamics
with stellar evolution, we have made a number of simplifying
assumptions about the cluster dynamics and populations. We attempted
to make these assumptions as reasonable as possible while still
retaining a tractable problem. In this section we attempt to
characterize the effects of the assumptions on our conclusions.

The first of the simplifying assumptions we have made is that the
density in the region of interest is constant. This is clearly not the
case in real globular clusters. The density decreases by a factor of
$\sim 2$ between the centre and the edge of the core in a King model. By
assuming that the core has a density equal to the central density, we
have overpredicted the number of blue stragglers.

We have also assumed that all populations have the same density at the
same position. Because of mass segregation, massive objects will be
more centrally concentrated, so the relative number density of objects
will vary radially. We expect that collisions involving binary stars
will happen more frequently closer to the centre of the cluster, and
that collisions involving more massive stars will happen closer to the
centre. Therefore, the shapes of the predicted distributions will vary with
radius in the cluster.

We have also assumed that the primary and secondary of each binary
system was drawn at random from an initial mass function. Therefore,
the mass function of the secondaries is independent of the primary
mass function. Observational (Abt 1990) and theoretical (Kroupa 1995)
evidence suggests that stars which form in aggregates of binary
systems with flat period distributions and component masses paired at
random will evolve dynamically to the field population seen
today. Therefore, our assumption of randomly paired masses is
appropriate for a young cluster, but we have neglected dynamical
modification of the binary mass ratios. During encounters involving
binary stars, higher mass secondaries are preferentially swapped into
the binary systems (Heggie 1975, Hut \& Bahcall 1983, Sigurdsson \&
Phinney 1995). Therefore, as the cluster evolves, the binary
population will evolve towards higher mass ratios. The collisions
involving these new binary stars will be more likely to produce more
massive collision products, and therefore the distribution of the blue
stragglers in the colour magnitude diagram will include more bright
blue stragglers. 

Since this paper concentrates on a method rather than detailed
comparison with observations, we chose cluster parameters which are
typical for globular clusters rather than trying to model M3 in
particular. For example, we have used a velocity dispersion of 10 km/s
in the core of the cluster. Observations (Gunn \& Griffin 1979)
suggest that a value of 6 km/s may be more realistic as a global
velocity dispersion in M3.  The velocity dispersion of the cluster
influences the distributions in two ways: it helps determine the flux
of stars, and it affects the cross sections for collisions involving
binary stars. The flux of stars is essentially a density multiplied by
the velocity dispersion. Since we have assumed a uniform velocity
dispersion for the entire region of interest, the velocity dispersion
is simply a constant factor which determines the total number of blue
stragglers.  As the velocity increases, more stars pass through a given
point each second, and more collisions occur. The dependence of the
cross sections on the velocity dispersion is rather more complicated.
In general, if the velocity of the stars decreases, the cross section
for each collision goes up since the stars are more likely to interact
gravitationally if they do not fly past each other quickly. The nature
of the collision can also change as the velocity of the stars changes,
since the orbits of the stars around each other are determined by the
stars' initial velocities.  Neglecting the effect of the velocity
dispersion on the cross sections, a value for $<v>$ of 6 km/s instead
of 10 km/s will reduce our total number of blue stragglers, bringing
our predictions closer to the observed number of bright blue
stragglers in the central 100'' of M3.

Since the velocity dispersion in the core of M3 is about 6 km/s,
binaries with longer periods can exist in M3 than in the theoretical
cluster with a larger velocity dispersion. The cutoff between hard and
soft binaries occurs at longer periods as the velocity dispersion
decreases. Soft binaries are disrupted early in the life of the
cluster. Therefore, the theoretical distributions in this paper have
included too little contribution from long period binaries than is
applicable for M3. As is demonstrated in figure 6, if we add long
period binaries to the distribution, the total number of stars is
increased, but the overall shape of the distribution is not greatly
affected.

We have also assumed that a single number, the velocity dispersion, is
sufficient to model the velocity distribution of stars in the
cluster. However, all stars are not moving with the same velocity, but
will have a nearly Maxwellian velocity distribution which is
characterized by the velocity dispersion. Therefore, we expect that
each quantity which is affected by the stellar velocity will have a
range of values. The flux of stars will follow a Maxwellian, and the
cross section for each collision will be an appropriately weighted
average of a range of cross sections. Both these effects should result
in smoothing of the theoretical blue straggler distributions. This
smoothing effect should be small for most globular clusters since the
velocity dispersion is fairly small.

The metallicity of M3 is [Fe/H]=-1.57 (Harris \& Racine 1979). The
evolutionary tracks used in this work were more metal-rich, with a
metallicity of [Fe/H]=-1.27 (Z=0.001). Therefore, all the evolutionary
tracks are too red and a little too faint for M3. However, since the
data are not calibrated to a standard system, we simply shifted the
evolutionary tracks by an amount which made the main sequence turnoff
match the turnoff for an M=0.8 M$_{\odot}$ star. Since any differential
shifts due to metallicity will be small, this change in metallicity
will simply shift the distributions without changing their shape or
the total number of predicted blue stragglers significantly.

\subsection{Two Populations of Blue Stragglers?}

Currently it is thought that two main mechanisms are responsible for
creating blue stragglers: direct stellar collisions, and the merger of
the two components of a binary system (Stryker 1993). In this work, we
have concentrated on the direct stellar collisions, but it is expected
that both mechanisms occur in different environments, perhaps even in
the same cluster. One piece of evidence for this has been the bimodal
radial distribution of blue stragglers in M3, and the different
luminosity functions of the inner and the outer populations (Bailyn \&
Pinsonneault 1995, but see also Sigurdsson \etal\ 1994). The results
of this paper suggest that two different mechanisms may be at work in
the same region of the cluster. We have shown that our theoretical
distributions have difficulty in simultaneously explain both the brightest blue
stragglers, and the width of the total distribution in the
colour-magnitude diagram. It seems likely that the brightest three
blue stragglers in the centre of M3 belong to a different population
than the rest of the blue stragglers. The globular cluster NGC 6397
shows a similar population of bright, central blue stragglers which may be
different from the rest of the blue stragglers in that cluster (Sills
\etal\ 1995). In both clusters, our theoretical distributions cannot
explain both the bright and the faint populations of blue stragglers
using a consistent set of assumptions.

The possible second population in M3 is brighter than the other blue
stragglers, and separated from the rest by a gap of about 0.5
magnitudes, and the stars in NGC 6397 have similar properties. It is
also separated from the horizontal branch by about the same amount. It
is unlikely that these stars are horizontal branch stars since the
horizontal branch is well defined. Therefore, we assume that these
stars are blue stragglers. Blue stragglers can be made bright in two
ways: they can have a large mass, or they can have a high helium
content. These blue stragglers could have a large mass because they
are the result of a collision between three stars instead of between
two. Triple collisions are expected to occur between three independent
stars (a very low probability occurrence), or to occur during an
encounter of a binary system with a single star, or during an
encounter between two binary systems.

We included the triple collisions which occurred during encounters
between binary and single stars our theoretical distributions.  The
structure of the collision products was determined using the same
algorithm as for the other `SPH' tracks. The evolutionary tracks of
some of these products do indeed pass through the region in the
colour-magnitude diagram in which these three stars exist. However,
the triple collisions included in these calculations do not contribute
significantly to the total number of blue stragglers (about 1\% for
case N). There are two reasons for this. First, the lifetimes of stars
with masses of about $2 M_{\odot}$ are much shorter than stars of $1.5
M_{\odot}$ by a factor of about 3. Second, the cross sections for
triple collisions are in general smaller than the cross sections for
collisions between two stars. This cross section is small because
triple collisions are difficult to cause when only three stars are
involved since there is no other star available to carry away angular
momentum from the system. The cross sections are also small because
triple collisions occur preferentially with small binary systems, and
the cross section is proportional to the binary semi-major axis.

In order to sustain the hypothesis that triple collisions are
responsible for the bright population, we need a larger contribution
of triple collisions to the blue straggler distributions.  It seems
likely that we can increase the cross sections for triple collisions
by considering encounters between two binary systems. We have not
included binary-binary interactions in this work since the cross
sections for the different kinds of products are not yet easily
calculated for a global population of binary stars, in the same way
that binary-single collisional cross sections can be calculated, using
STARLAB. Binary-binary interactions should produce many more collision
products for the same binary fraction, and the distribution of
collision products could be quite different. It is not yet clear that
the addition of binary-binary collisions can explain the bright
population of blue stragglers while maintaining the correct total
number of blue stragglers observed in clusters. Analysis of the cross
sections for binary-single and binary-binary encounters suggest that
unless the binary fraction is quite low, too many collisions take
place (Bacon \etal\ 1996).

The second method for creating bright blue stragglers is to increase
their helium content.  Blue stragglers can have a higher than expected
helium content if the star is made of two (or three) stars which
evolved significantly before the collision, and their central helium
is mixed to the surface of the star. The most extreme amount of mixing
allowed is represented by the fully mixed distributions presented in
this work. We have shown that stars which are fully mixed do indeed
become as bright and as blue as the three blue stragglers in
question. We can even produce these stars with the collisions between
two stars only, without having to invoke a third star, as long as the
product is fully mixed. Since blue stragglers are not fully mixed
during the collision or during the subsequent evolution, another
process must be at work if these stars are indeed fully or
significantly mixed. 

One possible process is rotation of the stars. We have neglected the
intrinsic rotation of the stellar collision products in this work
because we have been dealing with head-on collisions only. However,
when two stars collide in an off-axis collision, the resulting product
can be rotating as fast as 70\% of its break-up speed. Stars which
rotate quickly are subject to a number of instabilities which can
cause mixing of material, including meridional circulation (Tassoul
1978). Detailed modeling of the evolution of these rotating collision
products will be necessary to determine if these processes produce a
significant amount of mixing, enough to affect the position of the
stars in the colour-magnitude diagram (although see Ouellette \&
Pritchet 1998). Any investigation of the evolution of rotating blue
stragglers must also explain two seemingly incongruous facts: we
expect that stars of all masses may undergo rotational mixing,
suggesting a continuum of stellar properties, but yet the bright blue
stragglers in both M3 and NGC 6397 are separated by a gap from the
faint blue stragglers.

It is also possible that the merger of the two components of a binary
system can result in a star which is significantly mixed. This star
should be rotating rapidly also, so rotational mixing could play a
role. Also, if the merger involves any sort of common envelope
evolution, it is unclear whether the two cores of the stars remain
intact during the spiral-in phase or whether some significant portion
of the helium of these cores is mixed into the outer layer of the
star. If some mixing does occur, this may be enough to push some stars
into the region of the colour-magnitude diagram in question.  More
detailed models of the process of binary mergers is necessary before
this question can be answered.

\subsection{Future Work}

The above discussion of the two possible populations of blue
stragglers brings up a number of areas which still need to be
explored: encounters between two binary systems, rotation of the
non-head-on collision products, and binary mergers. We made a number
of other assumptions about the cluster which should be relaxed in
order to further understanding of the blue straggler phenomenon.

First of all, we have assumed that the inner region of this cluster is
a uniform box with the same density and velocity dispersion
everywhere. This is obviously not the case in real clusters. We should
use either a static (King) model or an evolving dynamical model of the
cluster to predict the density and velocity dispersion as a function
of radius, and apply those values to our calculations in such a way as
to take into account the stratification of the cluster.

We have neglected the evolution of the globular cluster in a number of
respects. We have assumed that the parents of these stellar collision
products are drawn from the current population of the cluster. We are
choosing stars which have evolved for 15 Gyr. However, collisions have
happened throughout the lifetime of the cluster, and so some of the
parents will have evolved for less time than we have assumed.
Collisions between stars with higher masses than the current turnoff
mass will also have occurred. We have also neglected to include the
effects of the collisions on the parent populations. For example, blue
stragglers can themselves be involved in stellar collisions sometime
after they were created. Also, interactions involving binary stars can
change the characteristics of the binary population. In order to model
the populations of collision products in a globular cluster properly,
we need to incorporate the detailed evolutionary tracks used in this
work into an N-body dynamical code, and follow the stellar collisions
in some Monte-Carlo fashion. 

Investigations of globular cluster dynamics with stellar collisions
has been begun by Portegies Zwart \etal\ (1997a,b). They have included
a simple prescription for normal stellar evolution in a detailed
N-body dynamics code, and have attempted to model the evolution of a
cluster in which stellar collisions occur. Their treatment of the
stellar evolution is relatively crude, consisting of a table of masses,
radii and luminosities. We have shown that the mass of the stellar
collision product alone is not enough to determine its evolutionary
state, and hence its effect on the cluster's evolution. However,
Portegies Zwart's treatment of the dynamics of the cluster is far more
detailed than anything we have attempted. We are approaching the same
problem from opposite directions, and our two approaches must
eventually be combined.

We should also expand this work to other globular clusters. This will
allow us to explore the effects of cluster metallicity and age on the
blue straggler population. We will also be able to make stronger
statements about the effects of cluster densities and central
concentrations on the populations in the cluster, and possibly comment
on the primordial binary population of the cluster. In order to draw
valid conclusions from theoretical predictions of a sample of
clusters, we should have high quality data to which we can compare our
models. These data should be photometrically complete down to the
turnoff of the cluster (or at least within a magnitude). They should
also be complete spatially, reaching into the core of the cluster
since most of the collisions will occur in the core. For many
clusters, particularly the dense (and hence interesting) ones, this
will require the Hubble Space Telescope since ground-based
observations of the cores of globular clusters are limited by crowding
effects. To be able to compare our theory with the data in a
statistically significant way, we also need on the order of 100 or
more blue stragglers in each cluster. As clusters are surveyed more
carefully and as the surveys reach into the cores, it is becoming
clear that M3 is not alone in its large population of blue stragglers
(e.g. 47 Tuc -- Edmonds 1998, M55 -- Mandushev \etal\ 1997). We feel
confident that in a few years, a sufficient data set will exist to
provide detailed comparison for our models.

\section{Summary}

We have described a method to model the distributions of blue
stragglers in the colour-magnitude diagram of globular clusters. By
combining the results of detailed stellar evolution models of blue
stragglers with relatively crude information about the dynamics of the
cluster, we can determine the total number of blue stragglers produced
by stellar collisions, and their expected positions in the
colour-magnitude diagram. As an example, we have applied this method
to the cluster M3, which has a large population of blue stragglers. We
assumed that the blue stragglers were created by stellar collisions,
and we restricted our investigation to the inner 100'' of the cluster.

The distribution of blue stragglers in the colour-magnitude diagram
and the total number produced by stellar collisions is determined by
the current dynamical state and population of the cluster, as well as
by the evolutionary paths of the blue stragglers themselves. We have
explored the effects of a number of different parameters on the shape
of the blue straggler distributions, and on the total number of blue
stragglers. We found that the binary fraction is the strongest factor
in determining the total number of blue stragglers in a cluster. The
mass distribution of the components of the binary system is more
important than the current stellar mass function in determining the
shape of the blue straggler distribution. The blue straggler
distribution is relatively insensitive to the period distribution of
the binary stars in the cluster.

We have also investigated the effect of the evolutionary history of
the blue stragglers on their distribution in the colour-magnitude
diagram.  Our theoretical distributions of blue stragglers cannot
simultaneously reproduce the brightest and the reddest blue stragglers
observed in M3.  We have determined that most of the blue stragglers
in the core of M3 can be produced by stellar collisions which did not
result in significant mixing of the stellar collision product. There
may also be a second population of brighter blue stragglers, which may
have undergone significant mixing during their creation. This
population consists of the brightest blue stragglers in M3, but may
also contain fainter stars.

In order for stellar collisions to produce the number of blue
stragglers observed in the inner regions of M3, a population of binary
stars must be present in the cluster. A binary fraction of 20\% is
sufficient to reproduce the observed numbers. The binary fraction and
the masses of binary stars seem to be more important in determining
the total numbers and distribution of blue stragglers than the
population of single stars. Small changes in the binary period
distribution do not affect the distribution of blue stragglers
significantly, but the total number of blue stragglers increases when
the period distribution favours longer periods.

To fully model collisional blue stragglers, this work can be and needs
to be expanded in a number of directions. We need to combine rigorous
stellar dynamics with our detailed evolution calculations. The theory
pertaining to blue stragglers needs to become more detailed.  And
finally, more detailed and more complete observations are required. We
have begun combining stellar evolution and stellar dynamics to model
globular clusters, but there is still much left to do, both
observationally and theoretically.

\acknowledgements This work was supported by NSF grant AST-9357387 and
NASA LTSA grant NAG5-6404. We would like to thank Steve McMillan for
the use of his STARLAB code. We would also like to thank P. Demarque,
F. Rasio and J. Lombardi for useful discussions.

\clearpage

\figcaption[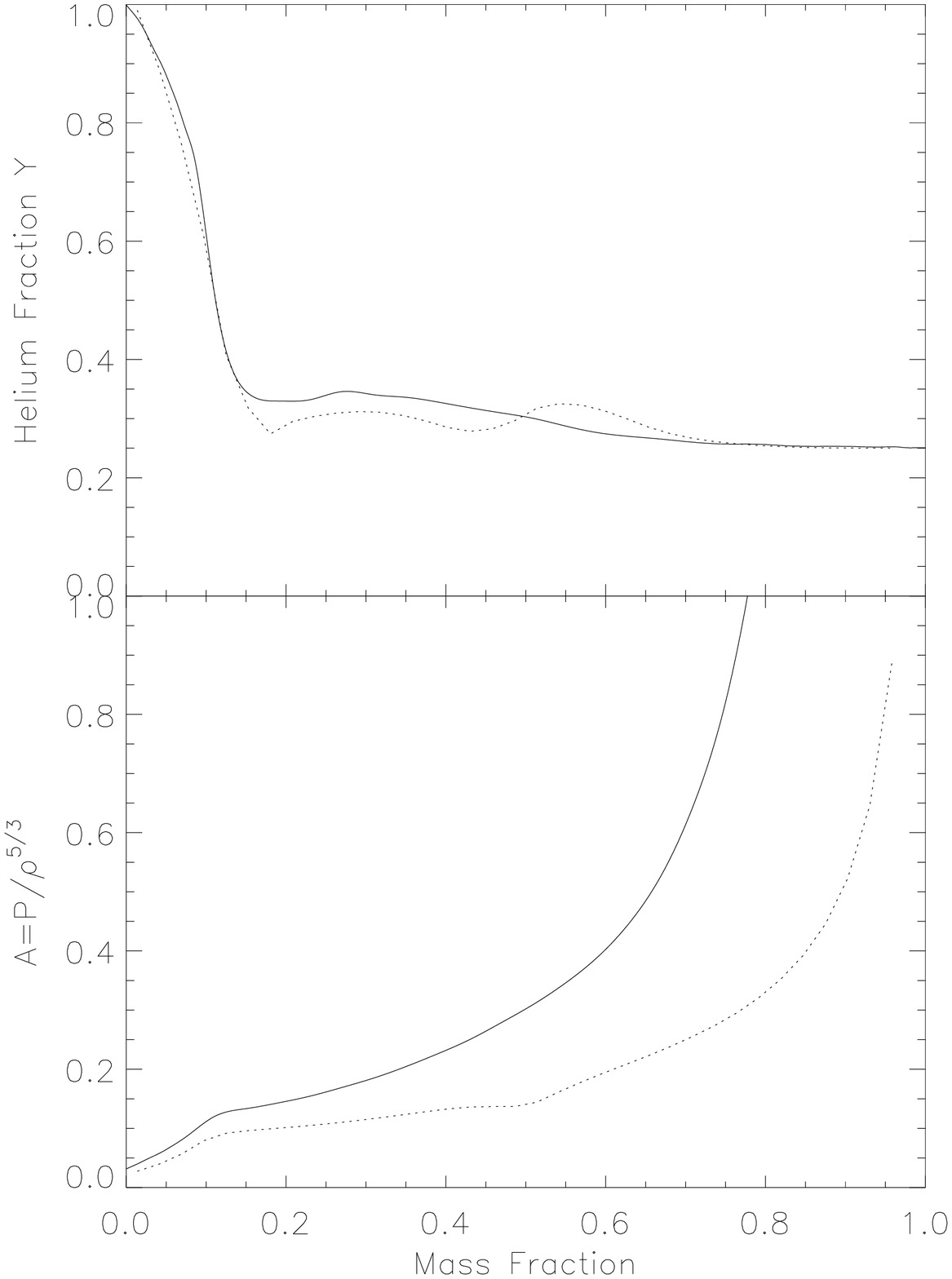]{Helium fraction and $A$ (entropy) profiles for a
collision between a $0.8 M_{\odot}$ star and a $0.4 M_{\odot}$
star. The solid line is the results of a SPH calculation of the
collision from Lombardi \etal\, and the dotted line was
determined from the algorithm described in this proposal. The helium
profile from the SPH calculation is smoother because of the tendency
of SPH codes to artificially smooth quantities over distances of
$\approx$3 smoothing lengths. The entropy profile of the SPH
calculation shows the effect of shock heating, which is more
significant in the outer parts of the collision remnant. }

\figcaption[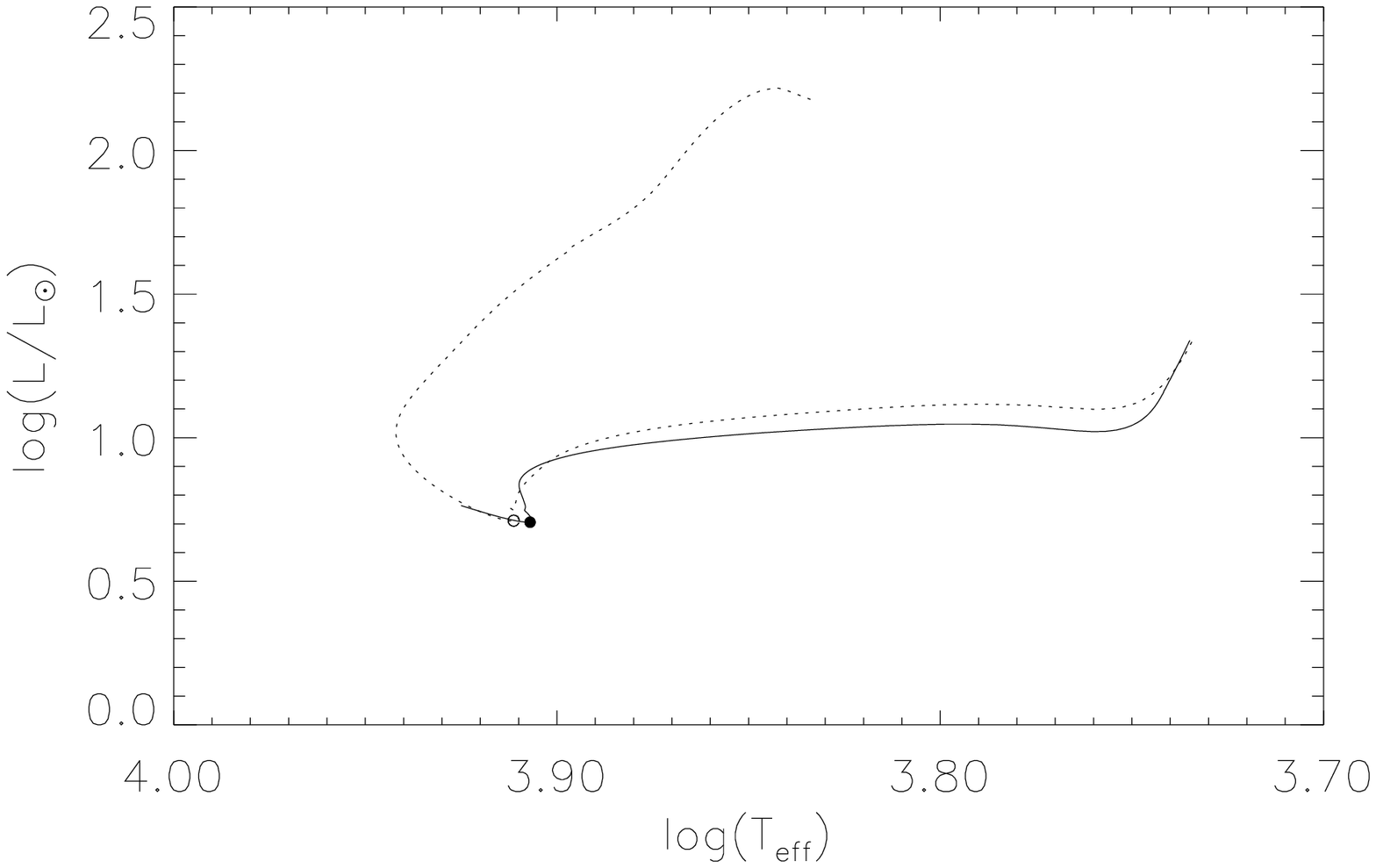]{Two evolutionary tracks for the product of a
collision between a 0.8 $M_{\odot}$ star and a 0.4 $M_{\odot}$
star. The evolution of the SPH calculation is given by the dotted
line. The initial conditions for the SPH calculation were given by
stellar models from YREC evolved to the turnoff age of the 0.8
$M_{\odot}$ star. The collision was followed using the SPH code, and
when the collision product reached hydrostatic equilibrium, it was
evolved using YREC through the thermal relaxation phase, to the main
sequence, and up the giant branch.}

\figcaption[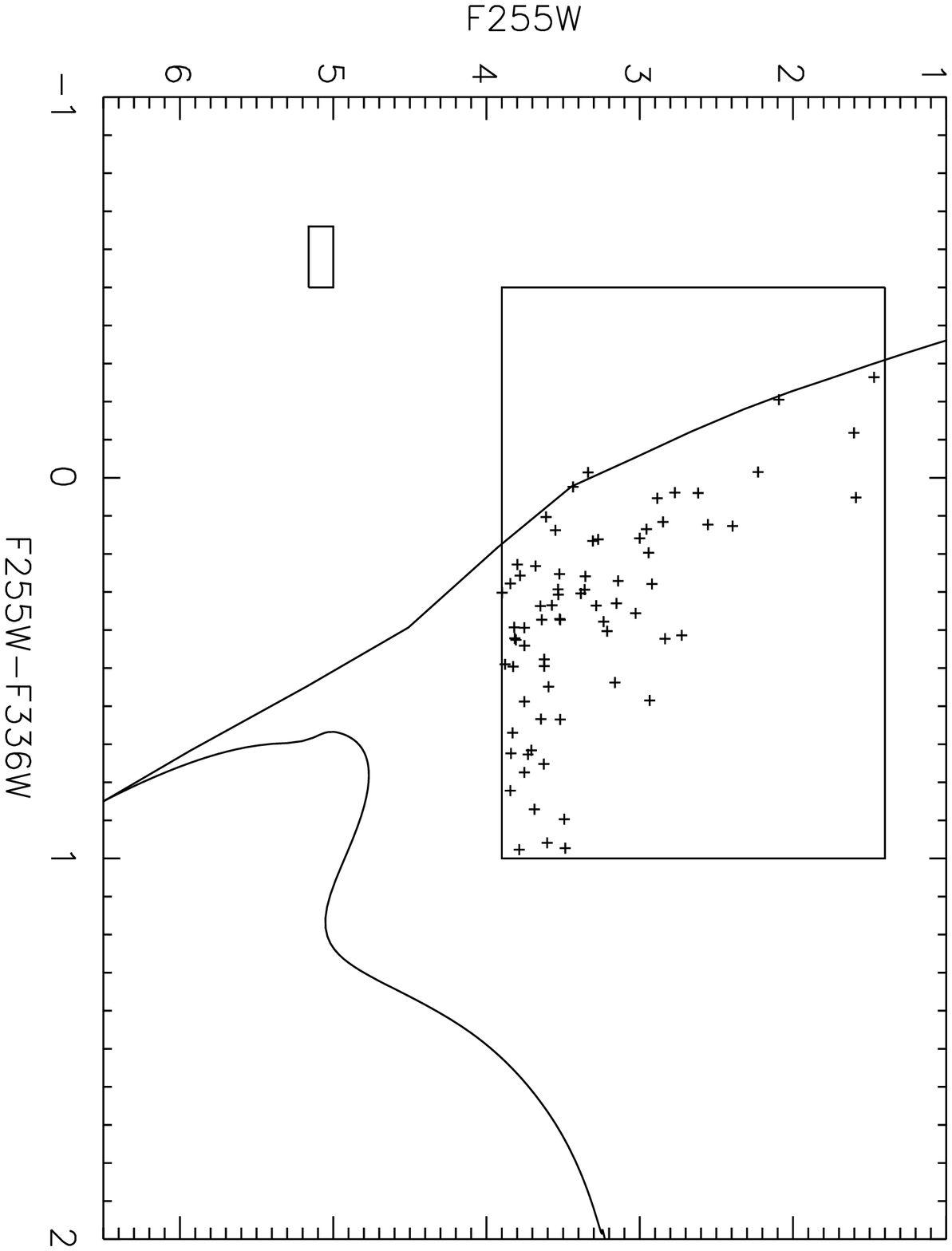]{This figure is designed to put the following
figures in perspective. This is a colour-magnitude diagram in the HST
ultraviolet filters F255W and F336W. The solid line on the left of the
diagram is the zero age main sequence. The other solid line is an
evolutionary track for a M=0.8M$_{\odot}$ star, which represents the
fiducial sequence of the cluster. The large box outlines the region of
the following four figures. The crosses are the blue straggler data
from Ferraro {\it et al} 1997. The density contours in the following
figures are at levels of 1, 5, 10 and 50 stars per box 0.16 magnitudes
on a side, which is represented by the small box in the bottom left
corner of this figure.}

\figcaption[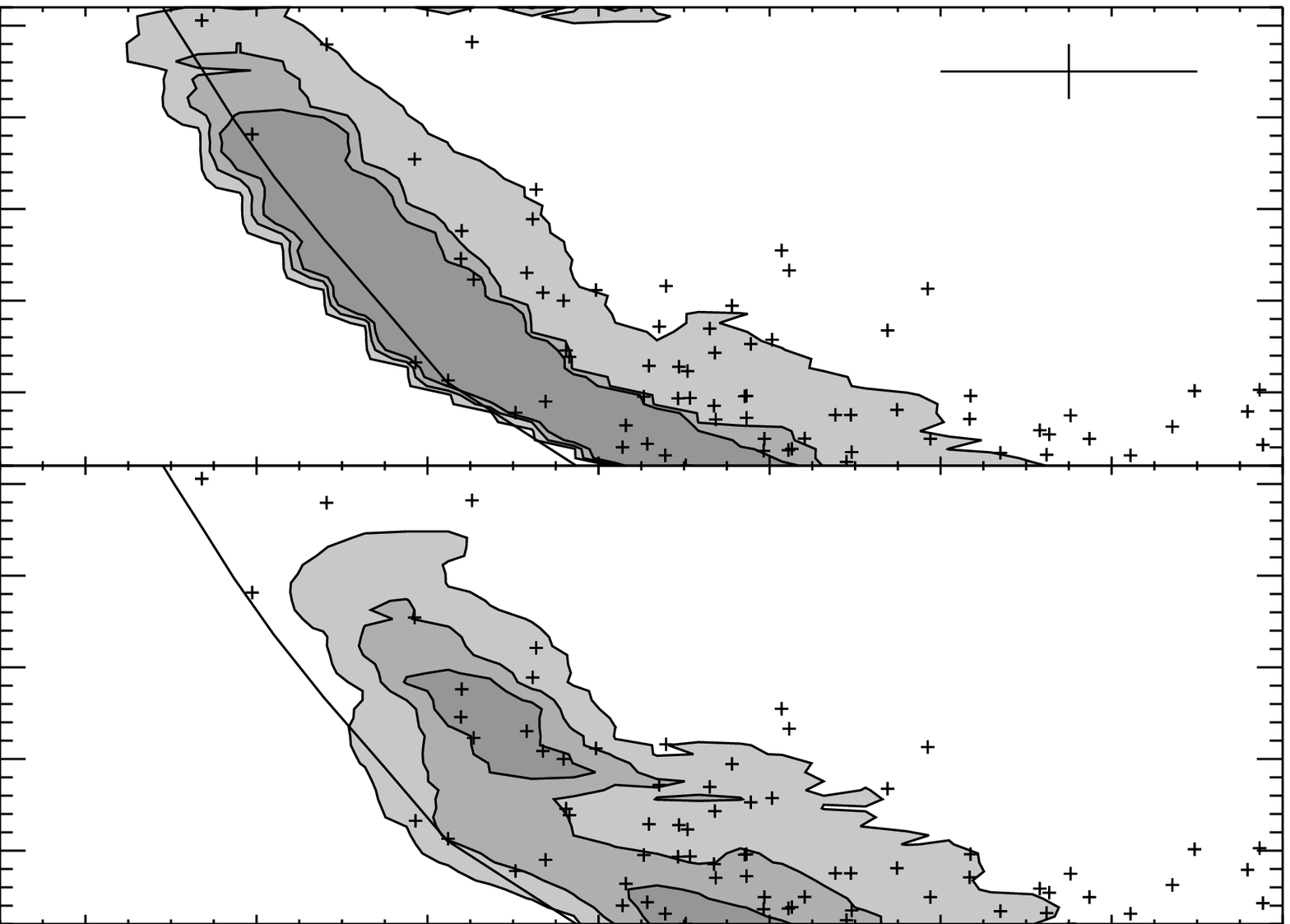]{These two distributions of blue stragglers in the
colour-magnitude diagram compare the effects of using different
evolutionary tracks for the blue stragglers. The top panel uses fully
mixed tracks, while the bottom panel uses the `SPH' tracks. Note the
differences in width and height between the two distributions, as well
as the difference in the total number of blue stragglers predicted.The
cross in the upper right corner gives the approximate photometric
error in the data at the main sequence turnoff. }

\figcaption[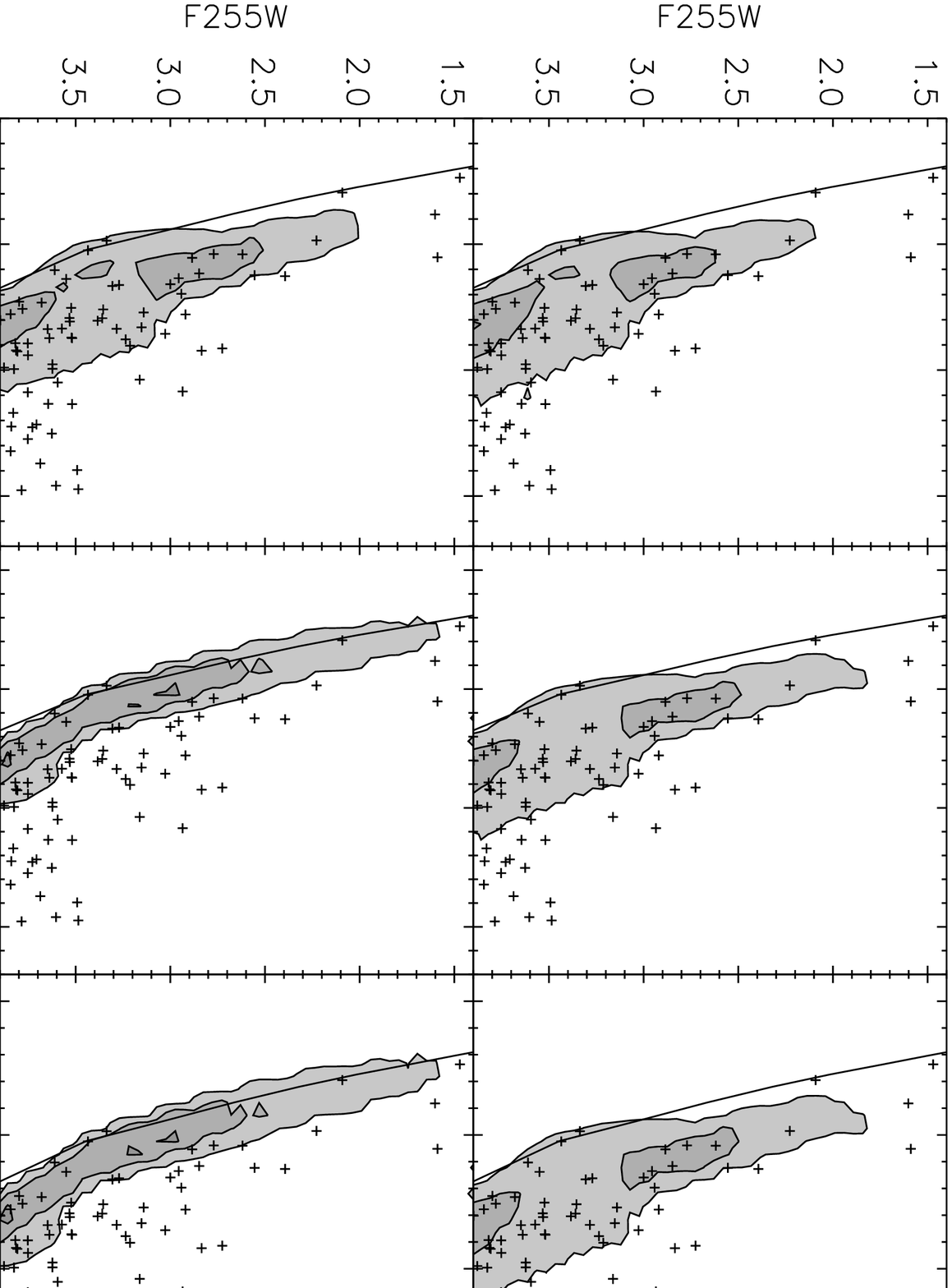]{These distributions demonstrate the effects
of changing binary fractions and evolutionary tracks. All these
distributions are for a binary component mass function with x=1.35, a
current mass function of x=-2, and a flat binary period
distribution. The distributions have been normalized to a total of 100
stars, to more clearly show the intrinsic differences in shape. The
top row of distributions all used the `SPH' tracks, while the bottom
row is for fully mixed tracks. The binary fractions for the three
columns are, left to right, 0\%, 20\% and 100\%.}

\figcaption[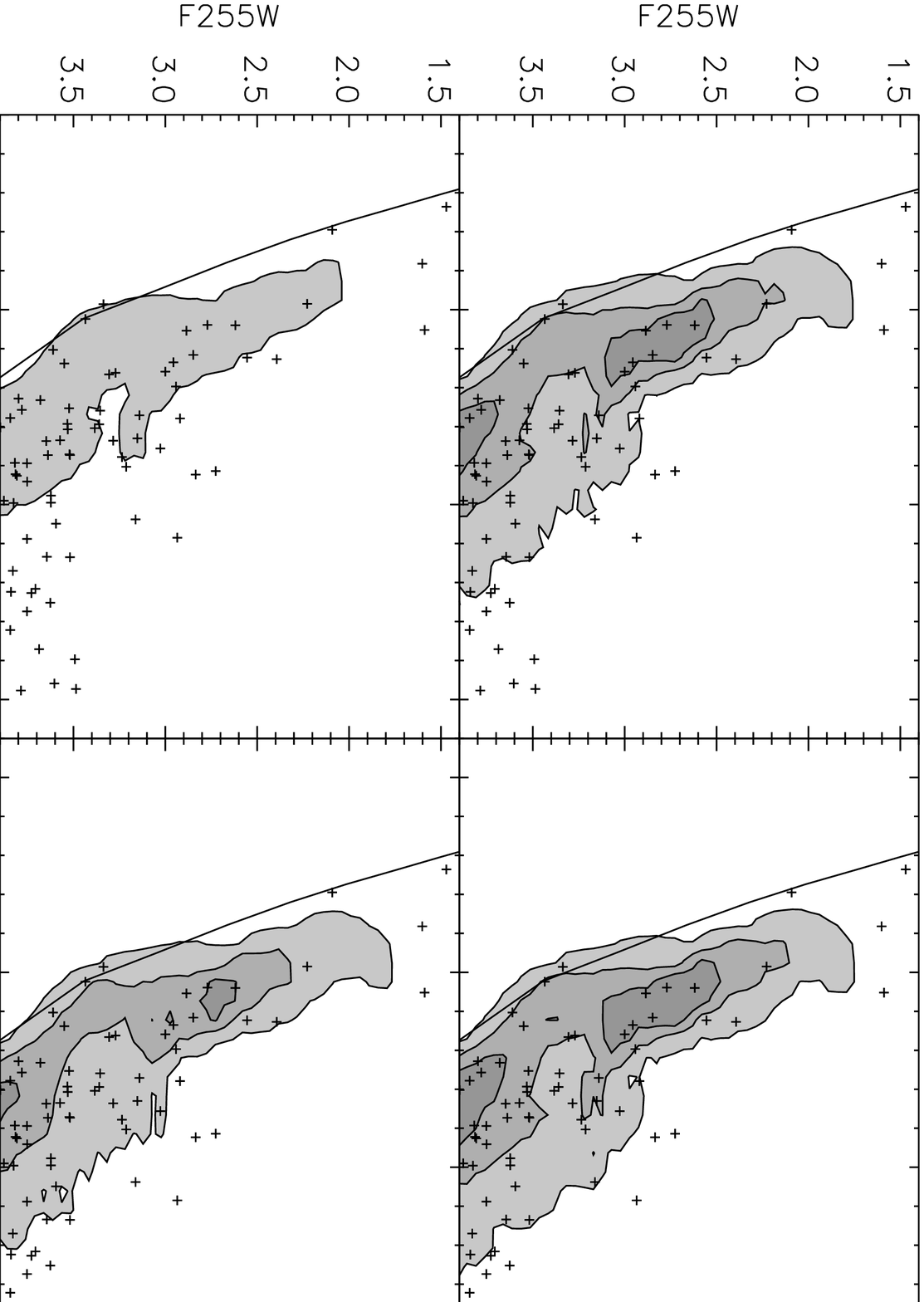]{These distributions demonstrate the effects of
changing the period distribution of the binary stars. The upper left
panel shows the distribution for periods which are evenly distributed
(flat) in $\log P$. The upper right panel shows the same distribution
but with the binary systems distributed according to the Duquennoy \&
Mayor (1991) period distribution. The bottom two panels show the flat
period distribution again, but with an artificial period cutoff. The
left-hand panel shows the contribution due to only those systems with
periods less than or equal to $1 \times 10^6$ seconds, while the
right-hand panel shows only those systems with periods greater than $1
\times 10^6$ seconds. These distributions have not been normalized,
and they are all for `SPH' tracks, a binary fraction of 20\%, a binary
component mass function with x=1.35, and a current mass function of
x=-2. }

\figcaption[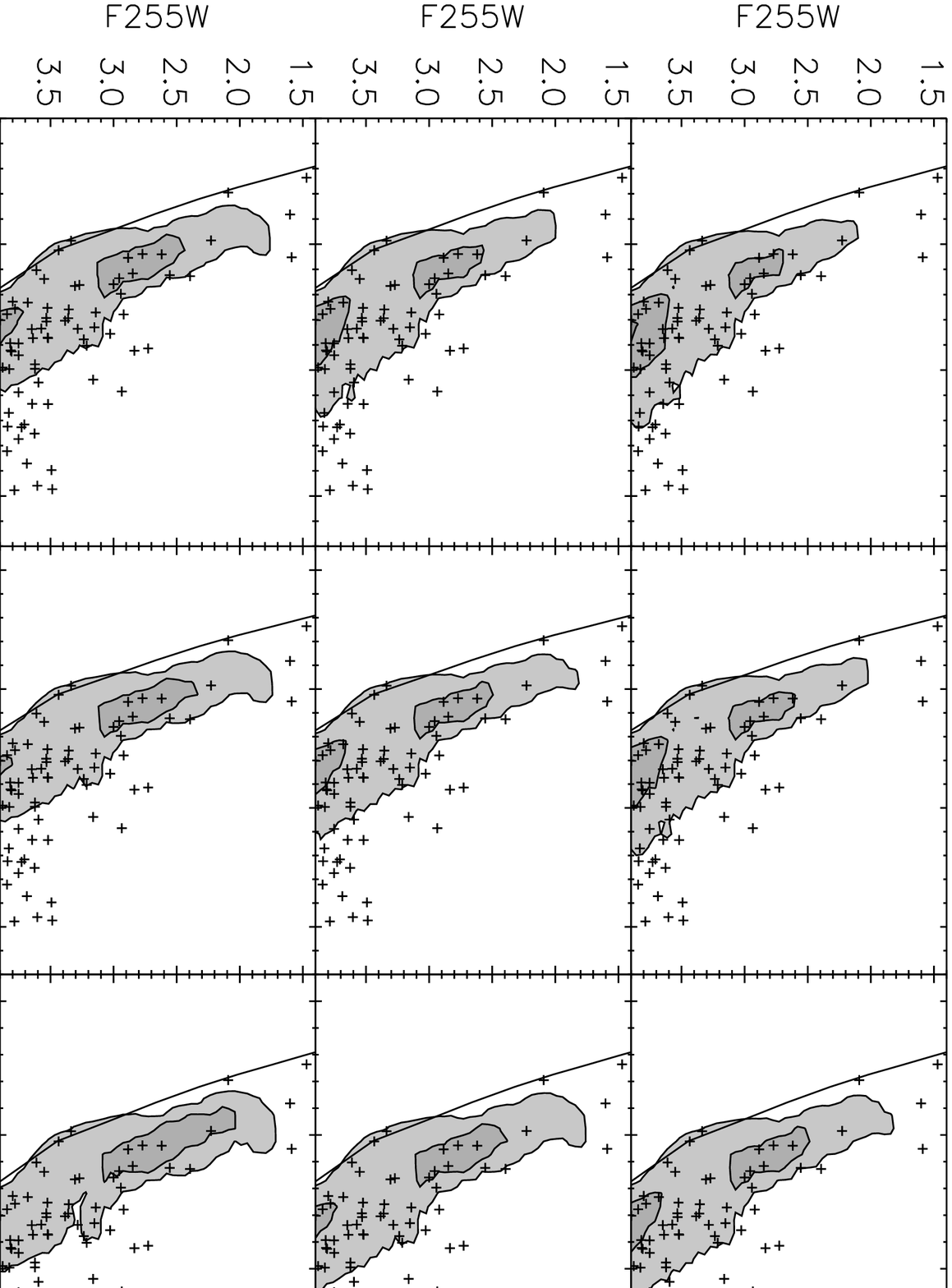]{This figure shows the effects of varying the
two mass functions used in this work. The current mass function in the
core of the cluster changes from $x=1.35$ to $x=0$ to $x=-2$ from left
to right across the diagram. The mass function from which the two
binary components were drawn changes from $x=1.35$ to $x=0$ to $x=-2$
from top to bottom of the diagram. Therefore, the top left
distribution is most biased towards lighter stars, while the bottom
right distribution is the most biased towards massive stars. All these
distributions have been normalized to 100 stars, use the `SPH' tracks,
have a binary fraction of 20\% and a flat period distribution.}

\figcaption[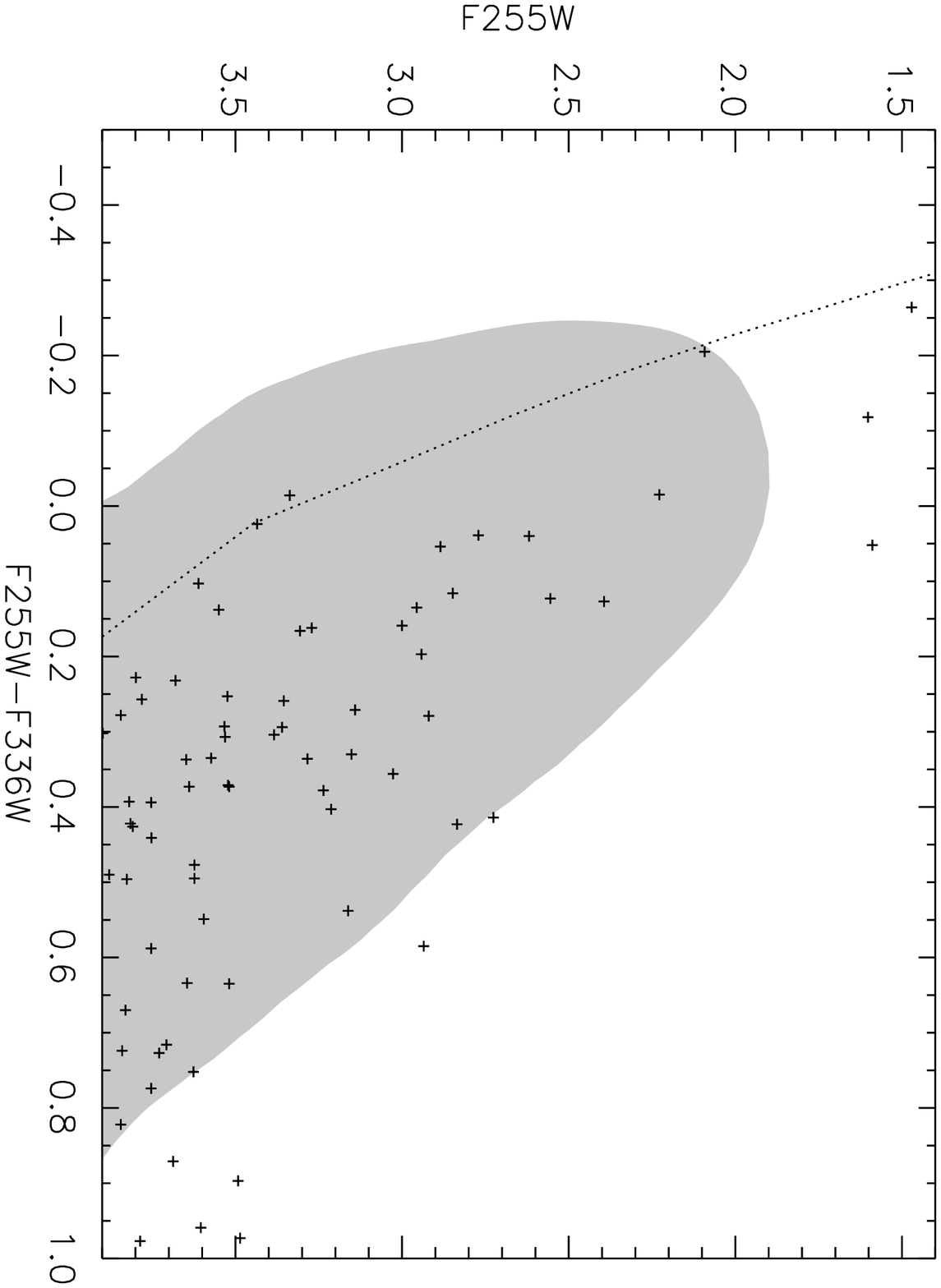]{In this figure we have taken our best fit to
the observational data (case J) and convolved it with a Gaussian of
width 0.15 in both colour and magnitude. The resulting distribution
indicates what effects the observational errors have on the
theoretical distributions.}

\clearpage
\plotone{fig1.eps}

\clearpage
\plotone{fig2.ps}

\clearpage
\plotone{fig3.eps}

\clearpage
\plotone{fig4.eps}

\clearpage
\plotone{fig5.eps}

\clearpage
\plotone{fig6.eps}

\clearpage
\plotone{fig7.eps}

\clearpage
\plotone{fig8.eps}

\clearpage
\begin{deluxetable}{ccccccc}
\tablecaption{Choices of parameters for the calculated distributions of blue stragglers in the colour-magnitude diagram.}
\tablehead{
\colhead{Case} & \colhead{Evolutionary} &\colhead{Binary} &\colhead{Period} &\colhead {$X_{current}$} & \colhead {$X_{binary}$} &\colhead{Total} \nl
 & \colhead{Tracks} &\colhead{Fraction} &\colhead{Distribution} & & &\colhead{Produced}
}
\startdata
A & SPH &   0\% &      & -2.00 &       &   24.7  \nl
B & FM  &   0\% &      & -2.00 &       &   53.3  \nl
C & SPH &   0\% &      &  1.35 &       &   16.8  \nl
D & SPH &   0\% &      &  0.00 &       &   20.9  \nl
E & SPH & 100\% & flat & -2.00 &  1.35 &  507.0  \nl
F & FM  & 100\% & flat & -2.00 &  1.35 & 1176.3  \nl
G & FM  &  20\% & flat & -2.00 &  1.35 &  427.5  \nl
H & SPH &  20\% & flat & -2.00 & -2.00 &  298.0  \nl
I & SPH &  20\% & flat & -2.00 &  0.00 &  224.5  \nl
J & SPH &  20\% & flat & -2.00 &  1.35 &  185.4  \nl
K & SPH &  20\% & flat &  0.00 & -2.00 &  245.9  \nl
L & SPH &  20\% & flat &  0.00 &  0.00 &  202.5  \nl
M & SPH &  20\% & flat &  0.00 &  1.35 &  178.7  \nl
N & SPH &  20\% & flat &  1.35 & -2.00 &  199.2  \nl
O & SPH &  20\% & flat &  1.35 &  0.00 &  174.2  \nl
P & SPH &  20\% & flat &  1.35 &  1.35 &  160.1  \nl
Q & SPH &  20\% & DM   & -2.00 & -2.00 &  331.8  \nl
R & SPH &  20\% & DM   & -2.00 &  0.00 &  248.8  \nl
S & SPH &  20\% & DM   & -2.00 &  1.35 &  204.5  \nl
T & SPH &  20\% & DM   &  0.00 & -2.00 &  272.8  \nl
U & SPH &  20\% & DM   &  0.00 &  0.00 &  223.5  \nl
V & SPH &  20\% & DM   &  0.00 &  1.35 &  196.1  \nl
W & SPH &  20\% & DM   &  1.35 & -2.00 &  220.4  \nl
X & SPH &  20\% & DM   &  1.35 &  0.00 &  191.5  \nl
Y & SPH &  20\% & DM   &  1.35 &  1.35 &  174.9  \nl
Z & SPH &  20\% & large& -2.00 &  1.35 &  149.0  \nl
AA& SPH &  20\% & small& -2.00 &  1.35 &   52.8  \nl
\enddata
\end{deluxetable}

\clearpage
\begin{deluxetable}{cccc}
\tablecaption{KS statistics: Comparison between theoretical distributions.}
\tablehead{
\colhead{Case} & \colhead{d} & \colhead {P (\%)} & \colhead{difference from case J}
}
\startdata
G & 0.34 & 8 $\times 10^{-8}$  & fully mixed tracks              \nl
A & 0.74 & 0.02                & binary fraction=0\%             \nl
E & 0.06 & 95                  & binary fraction=100\%           \nl
S & 0.04 & 99.9                & DM period distribution          \nl
Z & 0.08 & 90                  & $\log P > 6$                    \nl
AA& 0.35 & 1                   & $\log P < 6$                    \nl
H & 0.29 & 5 $\times 10^{-4} $ & binary mass function (x=-2)     \nl
P & 0.18 & 0.5                 & current mass function (x=1.35)  \nl
\enddata
\end{deluxetable}

\begin{deluxetable}{ccccc}
\tablecaption{KS statistics for selected cases.}
\tablecolumns{7}
\tablehead{
\colhead{Case} & \multicolumn{2}{c}{Luminosity Function} &\multicolumn{2}{c}{Temperature Function}  \nl
& \colhead{d} & \colhead{P(\%)} & \colhead{d} & \colhead{P(\%)}
}
\startdata
N & 0.250 & 63.2 & 0.375 & 16.2 \nl
J & 0.250 & 63.2 & 0.250 & 63.2 \nl
L & 0.312 & 34.8 & 0.375 & 16.2 \nl
R & 0.250 & 63.2 & 0.375 & 16.2 \nl
\enddata
\end{deluxetable}

\end{document}